\documentclass[preprint, showpacs,preprintnumbers,amsmath,amssymb,nofootinbib]{revtex4}
\usepackage{mathrsfs}
\usepackage{amsmath}
\usepackage[colorlinks, linkcolor=blue, citecolor=red, urlcolor=red]{hyperref}
\usepackage{amsfonts}
\usepackage{latexsym}
\usepackage{amsfonts}
\usepackage{graphicx}
\usepackage{epsf}
\usepackage{dcolumn}
\usepackage{bm}
\newcommand{\bea}[1]{\begin{eqnarray}\label{#1}}
 \newcommand{\eea}{\end{eqnarray}}

 \def\gsim{ \lower .75ex \hbox{$\sim$} \llap{\raise .27ex \hbox{$>$}} }
 \def\lsim{ \lower .75ex \hbox{$\sim$} \llap{\raise .27ex \hbox{$<$}} }
\def\/{\over}

\begin{document}

\title{\bf Interaction between two gravitationally polarizable objects induced by thermal bath of gravitons}

\author{Puxun Wu$^{1, 3}$, Jiawei Hu$^{1,2,}\footnote{Corresponding author at jwhu@hunnu.edu.cn}$ and Hongwei Yu$^{1,}\footnote{Corresponding author at hwyu@hunnu.edu.cn}$}
\address{ $^1$Department of Physics and Synergetic Innovation Center for Quantum Effects and Applications, Hunan Normal University, Changsha, Hunan 410081, China\\
$^2$Center for Nonlinear Science and Department of Physics, Ningbo
University,  Ningbo, Zhejiang 315211, China\\
$^3$Center for High Energy Physics, Peking University, Beijing 100080, China}

\begin{abstract}
The   quadrupole-quadrupole  interaction between a pair of gravitationally polarizable objects induced by vacuum fluctuations of the quantum linearized gravitational field is first obtained with a relatively  simple method, which is then used to investigate the contribution of thermal fluctuations of a bath of gravitons to the interaction at temperature $T$. Our result shows   that, in the high temperature limit, the contribution of thermal fluctuations   dominates over that of vacuum fluctuations and  the interaction potential behaves like $T/ r^{10} $, where $r$ is the separation between the objects, and in the low temperature limit, the contribution of thermal fluctuations is proportional to $T^{10}/r$, which only provides a small correction to the interaction induced by zero-point fluctuations. 

%{\bf keywords:} Gravitational interaction; linearized quantum gravity; vacuum fluctuations
\end{abstract}


\maketitle
\section{Introduction}

Fluctuating electromagnetic fields in vacuum  induce electric dipole moments in neutral atoms and thus generate between them the famous Casimir-Polder  (CP)  force which does not exist in classical electrodynamics~\cite{Casimir}.  The  CP force behaves like $r^{-7}$ in the retarded region and   $r^{-6}$ in the non-retarded region,  where $r$ is the separation between two atoms, and it has been extensively studied since its discovery~(for recent reviews, see Refs.~\cite{Lamoreaux05,Milton09,Bordag09}). In particular, it has  been generalized to include thermal corrections that arise from thermal fluctuations of a bath of thermal photons at nonzero temperature~\cite{Lifshitz}, where  it has been found that in the high temperature limit,  the  contribution of thermal fluctuations  dominates over that of the zero-point fluctuations  and gives rise to a  characteristic temperature ($T$) and distance dependence of   $T/r^{-6}$. In the low temperature limit, the thermal fluctuations only produce a small correction to the CP force dominated by  vacuum fluctuations. 

 Likewise, one would expect a  CP-like quantum correction to the classical Newtonian force between gravitationally polarizable objects due to vacuum fluctuations of gravitational fields when gravity is quantized. In this regard, the quantum correction has been computed between  a pair of gravitationally polarizable objects that arises from the induced quadrupole moments due to two-graviton exchange~\cite{Ford16} in close analogy with the calculation of the CP force between a pair of atoms from their induced dipole moments due to two-photon exchange~\cite{Sernelius},   and the quantum gravitational potential is found to behaves as $r^{-11}$ and $r^{-10}$ in the far and near regimes respectively.  It is also worth noting that the quantum gravitational correction between  two mass monopoles has been worked out by summing one-loop Feynman diagrams with off-shell gravitons in the framework where general relativity is treated as an effective field theory~\cite{Donoghue}.

Recently, based on the linearized quantum gravity and leading-order perturbation theory, we give an alternative derivation of this quantum correction due to  the vacuum fluctuation induced quadruple-quadruple interactions by treating two objects  as two-level harmonic oscillators from a quantum field theoretic prospect~\cite{Wu16}. This  approach is  parallel to that used by Casimir and Polder in studying  the quantum electromagnetic vacuum fluctuation induced electric dipole-dipole interaction between two neutral atoms in a quantum theory of electromagnetism~\cite{Casimir}.   Remarkably, this quantum correction can also been obtained by a calculation of the scattering amplitude of two-graviton exchange~\cite{Holstein2016}. 

As a natural step forward, in this paper, we plan to investigate what happens to the gravitational interaction  between  gravitationally polarizable objects when they are in a thermal bath of gravitons rather than in a vacuum. Let us note that 
a thermal bath of gravitons may be created by the Hawking effect or cosmological particle production.

In this paper, we first present another different but rather simple  approach to derivate  the quadruple-quadruple interaction  between a pair of gravitationally polarizable objects. Then, we consider the contribution of thermal fluctuations by immersing two objects in a thermal bath.  Throughout this paper, the Latin indices run from $0$ to $3$, while the Greek letter is from $1$ to $3$.  The Einstein convention is assumed for repeated index and $\hbar=c=k_{B}=1$ is set. Here, $\hbar$ is the reduced Planck constant  and $k_{B}$ is the Boltzmann constant. 

\section{Simple derivation of the quadruple-quadruple  interaction} \label{secfieldequ}
We assume that there is a pair of gravitationally polarizable objects, labeled $A$ and  $B$, at ${\bf r}_A$ and ${\bf r}_B$ with respect to an arbitrary origin and  the metric can be expressed as  $g_{\mu\nu}=\eta_{\mu\nu}+h_{\mu\nu}$, where $h_{\mu\nu}$ describes  the  fluctuating vacuum gravitational fields which are quantized~\cite{Yu99}. A direct consequence of quantization of gravity is the lightcone fluctuations which have been examined in~\cite{Yu99, Yu00, Yu03, Yu09}. In the present paper, we are concerned with  the correction they induce to the classical Newtonian interactions.

At an arbitrary point ${\bf r}$, the vacuum field of linearized gravity has the standard form~\cite{Oniga} 
\bea{hij}
h_{ij}({\bf r})&=&\sum_{{\bf k}, \lambda} \sqrt{\frac{8\pi G}{ (2\pi)^{3}\omega}} \big [a_{\lambda}(\omega) e_{ij}({\bf k}, \lambda) e^{i {\bf k} \cdot {\bf r}-i \omega t}+H. c. \big ] \nonumber \\
&\equiv& \sum_{{\bf k}}h^{\bf k}_{ ij}({\bf r})\;,
\eea
where the transverse tracefree (TT) gauge is taken, $H.c.$ denotes the Hermitian conjugate, $a_{\lambda}(\omega) $ is the gravitational field operator,  which defines the vacuum  $a_{\lambda}(\omega) | \{0\}\rangle=0$,  $\lambda$ labels the polarization states, $e_{ij}({\bf k}, \lambda)$ are polarization tensors,  $\omega=|{\bf k}|=(k_{x}^{2}+ k_{y}^{2} +k_{z}^{2})^{1/2}$, and $G$ is the Newton's gravitational constant.

The fluctuating vacuum fields induce quadrupole moments in  the objects $A$ and  $B$, which are written as 
\bea{Po}
Q_{i, lm}(\omega)=\alpha_i(\omega)E^{\bf k}_{lm}({\bf r}_i)\,, \quad i=A \; {\text or}\;  B\;,
\eea
where $\alpha_i(\omega)$ represents  the  polarizability of object $i$, which we assume to be  isotropic for simplicity,   and $E_{lm}=\sum_{\bf k}E^{\bf k}_{lm} $ is the gravito-electric tensor   defined  as 
\bea{Emn} E_{lm}=-R_{0l0m}\eea by an analogy between the linearized Einstein field equations and the Maxwell equations~\cite{Campbell}. Here,  $R_{\mu\nu\alpha\beta}$ is  the Riemann tensor. 
From the definition of $E_{lm}$,  we have 
\bea{Eij}
E_{lm}=\frac{1}{2}\ddot{h}_{lm}\;,
\eea
where a dot denotes a derivative with respect to time $t$.

If the object $A$ is polarized by vacuum fluctuations, it will emit  the gravitational waves, which  generate a field at the position of object $B$: $E^{\bf k}_{ij}(A\rightarrow B)$. 
  As a result, there exists an inevitable interaction between the quadrupole moment of object $B$ induced by vacuum fluctuations and the field  ${E}^{\bf k}_{ij} (A\rightarrow B)$.  
  The interaction potential   can be obtained perturbatively.  It is well known that the energy of a localized mass distribution $\rho(x)$ in the presence of an external potential $\Phi(x)$ is
\begin{equation}
V_{t}=\int \rho(x)\Phi(x)d^3x\;.
\end{equation}
If we assume that the potential $\Phi(x)$ varies slowly over the region where the mass is located, then the external potential can be expanded in a Taylor series as
\begin{equation}
\Phi(x)=\Phi(x_0)+x_i \frac{\partial \Phi(x_0)}{\partial x_i}
    +\frac{1}{2}x_ix_j\frac{\partial^2\Phi(x_0)}{\partial x_i\partial x_j}+\cdots\;.
\end{equation}
Because there is no mass dipole  in gravitation, the leading term of the interaction between two objects after the Newtonian potential is the quadrupole term
\begin{equation}
V=\frac{1}{2}\int d^3x \rho(x)x_ix_j 
   \frac{\partial^2\Phi}{\partial x_i\partial x_j}\;.
\end{equation}
Note that $\nabla^2\Phi=0$ in an empty space, so we can rewrite the above equation  as
\begin{equation}\label{V}
V=-\frac{1}{2}Q_{ij}E_{ij}\;,
\end{equation}
where
\begin{equation}
Q_{ij}=\int d^3x \rho(x)\left(x_ix_j -\frac{1}{3}\delta_{ij}r^2 \right)\;,
\end{equation}
and
\begin{equation}\label{E_ij}
E_{ij}=-\frac{\partial^2\Phi}{\partial x_i\partial x_j}
       +\frac{1}{3}\delta_{ij}\nabla^2\Phi\;.
\end{equation}
In general relativity,  $E_{ij}$ is defined in Eq.~(\ref{Emn}), which can be shown to coincide with the expression given in Eq. (\ref{E_ij}) in the Newtonian limit.

Thus, from Eq.~(\ref{V}), one can see that the potential from quadrupole-quadrupole  interaction between objects $A$ and $B$  takes the form  
\bea{HA}
V(r)=-\frac{1}{2} \sum_{\bf k} \langle \{0\} |   {Q}_{B, ij}(\omega)  {E}^{\bf k}_{ij} (A\rightarrow B)|\{0\} \rangle \; ,
\eea
where  $r=|{\bf r}_{A}-{\bf r}_{B}|$ is the separation between objects $A$ and $B$. At the same time, the interaction potential can also be expressed as 
\bea{HB}
V(r)=-\frac{1}{2} \sum_{\bf k} \langle \{0\} |   {Q}_{A, ij}(\omega)  {E}^{\bf k}_{ij} (B\rightarrow A)| \{0\} \rangle \eea
if we exchange roles of  objects $A$ and  $B$ in the above physical picture.  Using Eq.~(\ref{Po}), the above expression can be rewritten as 
\bea{HA3}
V(r)=-\frac{1}{2 }   \sum_{\bf k}  \langle 0 | \alpha_A(\omega) E^{\bf k}_{A, ij} ({\bf r}_A)  {E}^{\bf k}_{ij} (B\rightarrow A) |0 \rangle \;.
\eea
With $\hat{{\bf n}}={\bf r}/r $, the field ${E}^{\bf k}_{ij} (B\rightarrow A)$ is given by~\cite{Ford16}
\bea{wee}
{E}^{\bf k}_{ij} (B\rightarrow A) =  \mathrm{Re} \bigg\{\frac{G e^{ir\omega}}{r^5} \Lambda^{lm}_{ij}(\omega r, \hat{{\bf n}}) \, Q_{B,lm} (\omega) \bigg\}\ ,
\eea
where 
\bea{Lam}
 \Lambda^{lm}_{ij}(\gamma, \hat{\bf n}) &=&\frac{1}{2} [(6-6 \gamma^2+2 \gamma^4- 6 i \gamma+ 4 i \gamma^3)\delta^l_i \delta^m_j \nonumber \\
 &&+(-15+9 \gamma^2-\gamma^4+15 i \gamma- 4 i \gamma^3) (n^in^l\delta^m_j+n^jn^l\delta^m_i+ n^in^m\delta^l_j+n^jn^m\delta^l_i)\nonumber \\
 && +(-15+3 \gamma^2 + \gamma^4+15 i \gamma + 2 i \gamma^3)n^l n^m\delta_{ij} \nonumber \\
 && +(105- 45 \gamma^2 +\gamma^4- 105 i \gamma + 10 i \gamma^3)n^in^jn^l n^m]\ .
\eea

Since $Q_{B,lm} (\omega)$ is induced by vacuum fluctuations, it can also be expressed as $Q_{B, lm}(\omega)=\alpha_B(\omega)E^{\bf k}_{B, lm}({\bf r}_B)$. Thus, to obtain  the interaction potential given in Eq.~(\ref{HA3}), we need to calculate the following vacuum expectation value
\bea{vmv}
 \langle \{ 0\} |  E^{\bf k}_{A, ij}({\bf r}_A)  {E}^{\bf k}_{B, lm} ({\bf r}_B) | \{0\} \rangle &=& \frac{\omega^4}{4}\langle \{0\} | {h}^{\bf k}_{ij}( {\bf r}_A) {h}^{\bf k}_{lm}( {\bf r}_B) |\{0 \} \rangle \nonumber \\
&=&   \frac{ G\omega^3}{ (2\pi)^{2}}  \sum_{\lambda} e_{ij}({\bf k}, \lambda) e_{lm}({\bf k}, \lambda)  e^{i{\bf k}\cdot { ({\bf r}_A-{\bf r}_B)}} \nonumber \\
&=&    \frac{ G\omega^3}{ (2\pi)^{2}}  g_{ijlm} (\hat{{\bf k}}) e^{i{\bf k}\cdot { \bf r}} \ ,\eea
where $\hat{{\bf k}}={\bf k}/\omega$, and  the summation of polarization state gives~\cite{Yu99}
 \bea{eijekl}
g_{ijlm} (\hat{{\bf k}})  &=&\delta_{il}\delta_{jm}
+\delta_{im}\delta_{jl}-\delta_{ij}\delta_{lm}
+\hat k_i\hat k_j \hat k_l\hat k_m+\hat k_i \hat k_j \delta_{lm} \nonumber\\
&&+\hat k_l \hat k_m \delta_{ij}-\hat k_i \hat k_m \delta_{jl}
-\hat k_i \hat k_l \delta_{jm}-\hat k_j \hat k_m \delta_{il}-\hat k_j \hat k_l \delta_{im}\,.
\eea
Substituting Eqs.~(\ref{wee}, \ref{vmv}) into Eq.~(\ref{HA3}) leads to
 \bea{HA6}
V(r)=-\frac{G^{2}}{8\pi^2 } {\mathrm Re} \bigg\{ \sum_{\bf k}   \alpha_A(\omega)  \alpha_B(\omega) \frac{\omega^3 e^{i({\bf k}\cdot {\bf r}+ \omega r)}}{r^5} \Lambda^{lm}_{ij}(\omega r, \hat{{\bf n}}) \,   g_{ijlm}(\hat{\bf k}) \bigg\} \;.
\eea
Using $\hat{\bf k} \cdot \hat{\bf n}=\cos(\theta)$, we obtain
\bea{}
 \Lambda^{lm}_{ij}(\gamma, \hat{{\bf n}}) \,   g_{ijlm}(\hat{\bf k})& =& \frac{1}{16}\big [-27 + 27 i \gamma + 39 \gamma^2 - 30 i \gamma^3 - 35 \gamma^4 \nonumber \\
 &&-   4 (15 - 15 i \gamma - 27 \gamma^2 + 22 i \gamma^3 + 7 \gamma^4) \cos(2\theta) \nonumber  \\
    && - (105 - 105i \gamma - 45 \gamma^2 + 10 i \gamma^3 + \gamma^4) \cos(4 \theta) \big]
     \eea
  Now we replace  the summation   in Eq.~(\ref{HA6}) by  integral  and then change into the spherical coordinate
     \bea{}
      \sum_{\bf k}  \rightarrow \int d^{3}{\bf k}  \rightarrow \int_0^\infty k^2 dk \int_0^\pi d\theta\, 2\pi \ .
      \eea
Making a further replacement $e^{i{\bf k}\cdot {\bf r}}=e^{i \omega r \cos(\theta)} \rightarrow \cos(wr \cos(\theta))$ and taking the real part in Eq.~(\ref{HA6}), we obtain 
\bea{HA7}
V(r)=-\frac{G^{2}}{\pi r^{10} } \lim_{\eta \rightarrow 0} \int_0^\infty d \omega\  e^{-\eta \omega}  \alpha_A(\omega)  \alpha_B(\omega) f(\omega r) \;,
\eea
where a convergence factor is introduced to avoid  divergence and 
\bea{}
f(\gamma) &=& (-630 \gamma + 330 \gamma^3 - 42 \gamma^5 + 4 \gamma^7) \cos(2\gamma) \nonumber\\
&& + (315 - 585 \gamma^2 + 129 \gamma^4 - 14 \gamma^6 + \gamma^8) \sin(2\gamma)\ .
\eea
Assuming an approximate static polarizability $\alpha_{A(B)}(0)$ and performing the integration in   Eq.~(\ref{HA7}), one can obtain 
\bea{z0}
 V(r)=-\frac{3987G^{2}}{4\pi r^{11} }\alpha_A(0)  \alpha_B(0)  \; ,
 \eea
which is the same as what was obtained in~\cite{Ford16, Wu16, Holstein2016} with different methods.

\section{the interaction from thermal fluctuations}

Now we begin to study the  contribution from thermal fluctuations with the method we just introduced in the proceeding section. We assume that two objects are in a thermal bath of gravitons at a temperature $T$. Then, the interaction potential takes the form 
\bea{HAT1}
V(r)=-\frac{1}{2 }   \sum_{\bf k}  \langle \{\beta\} | \alpha_A(\omega) E^{\bf k}_{A, ij}  {E}^{\bf k}_{ij} (B\rightarrow A) |\{\beta\} \rangle \; ,
\eea
where $\beta=1/T$. For a thermal state of gravitons $ |\{\beta\} \rangle $,  one has the following relations 
\bea{}
\langle \{\beta\} | a^{\dagger}_{\lambda}(\omega) a_{\lambda}(\omega) |\{\beta\}\rangle= N(\beta, \omega)\ , \quad 
\langle \{\beta\} | a_{\lambda}(\omega) a^{\dagger}_{\lambda}(\omega) |\{\beta\}\rangle=1+ N(\beta, \omega) \ ,
\eea
 where
\bea{}
N(\omega, \beta)=\frac{1}{e^{\beta \omega}-1} \ .
\eea
Following the same procedure as in the proceeding  section, one can obtain 
\bea{HAT2}
V(r)&=&-\frac{G^{2}}{\pi r^{10} } \lim_{\eta \rightarrow 0} \int_0^\infty d \omega\  e^{-\eta \omega}  \alpha_A(\omega)  \alpha_B(\omega) f(\omega r) [1+2N(\beta, \omega)] \nonumber \\
&\equiv& V_{0}(r)+V_{T}(r)\;,
\eea
where $V_{0}(r)$ and $V_{T}(r)$ represent the contributions from vacuum fluctuations and thermal fluctuations, respectively. 

Taking the high-temperature  limit,  which means that $\beta \ll r $, we find that
\bea{T1}
V_{T}(r)=-\frac{ 315 G^{2}}{\beta r^{10} }\alpha_A(0)  \alpha_B(0) 
\eea
Since the interaction  potential is determined by both the zero-point fluctuations and the thermal ones, we have
\bea{}
 V(r)=-\frac{G^{2}}{r^{11} }\alpha_A(0)  \alpha_B(0) \bigg[ \frac{3987}{4\pi  }+ \frac{ 315 r}{\beta  } \bigg]\; .
 \eea
 It is easy to see that since   $\beta \ll r $  the thermal fluctuations play a dominating contribution on the potential and this potential is proportional to $r^{-10} \beta^{-1}$. 
 
In  the   low-temperature limit  ($\beta \gg r $),  one has
\bea{T2}
V_{T}(r)=-\frac{83456 \pi^9 G^{2}}{10395} \frac{1} {r \beta^{10}} \alpha_{A}(0)\alpha_{B}(0)\ .
\eea
From Eqs.~(\ref{z0}) and (\ref{T2}), the potential is 
\bea{}
 V(r)=-\frac{G^{2}}{r^{11} }\alpha_A(0)  \alpha_B(0) \bigg[ \frac{3987}{4\pi  }+\frac{83456 \pi^9 }{10395} \frac{r^{10}} { \beta^{10}} \bigg]\; ,
 \eea
The contribution from thermal fluctuations is proportional to $r^{-1}\beta^{-10}$, which provides a correction term to the potential dominated by  the vacuum fluctuations.  

Now a question arises as to when the thermal correction might become appreciable as compared to the classical Newtonian interaction in the high temperature limit. 
To address this issue, let us first consider a simple system of mass, i.e., an elastic sphere.
If the object has  radius $R_{i}$, mass $M_{i}$ and frequency $\omega_{i}$, we have $\alpha_{i}\sim M_{i}R_{i}^{2}/\omega_{i}^{2}$  from the dimensional analysis. For convenience,  we introduce the orbital frequency $\Omega_{i}=\sqrt{GM_{i}/R_{i}^{3}}$ of a gravitationally bound system as the reference frequency, which sets a lower bound on $\omega_{i}$ for any physical system.  Then, in the high temperature limit,  we have
 \bea{}
 V(r)\simeq -{315T}\bigg(\frac{\Omega_A \Omega_B}{\omega_A \omega_B}\bigg)^{2}  \frac{R_{A}^{5}R_{B}^{5}}{r^{10}}\; .
 \eea
Assuming that $R_{A}=R_{B}=R$, and $M_{A}=M_{B}=M$ for simplicity, we can obtain the ratio between $V(r)$ and the classical Newtonian gravitational potential   $V_{N}(r)=-GM^{2}/r$ as
\bea{33}
\frac{V(r)}{V_{N}(r)}=315\frac{k_{B}T}{{GM^2}/{R}}
 \bigg(\frac{\Omega}{\omega_{0}}\bigg)^{4} \bigg(\frac{R}{r}\bigg)^{9}\ ,
\eea
where $\omega_{0}=\omega_{A}=\omega_{B}$, $\Omega=\Omega_{A}=\Omega_{B}$ and  we returned to the SI units. 
For a gravitationally bounded object, we have $\Omega_{i} \sim \omega_{i}$. Thus, if the temperature is high enough ($k_{B}T\approx \frac{GM^2}{R}$), the interaction between two massive objects, which results from thermal fluctuations of gravitons, may  become appreciable to the classical Newtonian interaction.  However, for a gravitationally bounded object, the gravitational potential energy  is of the order of $-GM^2/R$. When the temperature of the environment  grows high enough ($k_B T\sim GM^2/R$), such gravitationally bounded objects will break up. Therefore, for a gravitationally bounded object, we have 
$\frac{V(r)}{V_{N}(r)}\sim \left(\frac{R}{r}\right)^{9}$ for the best. For an electrically bounded object, the breakup  temperature  is higher, i.e. $k_B T \sim \hbar\omega_0$. However, since $\omega_0\gg\Omega$ for an electrically bounded object, the ratio (\ref{33}) is suppressed. At the temperature $k_B T=\hbar\omega_0$, Eq. (\ref{33}) can be rewritten as 
\bea{34}
\frac{V(r)}{V_{N}(r)}
=315\left(\frac{l_P}{R}\right)^2  \left(\frac{c}{\omega_0 R}\right)^3
    \bigg(\frac{R}{r}\bigg)^{9}\ ,
\eea
where $l_P=\sqrt{\hbar G/c^3}$ is the Planck length.  The radius $R$ is typically  much greater than the Planck length. As a result, the ratio (\ref{34}) is usually extremely  small, although the transition wavelength $c/\omega_0$ may be large compared to $R$. As a concrete example, we consider the interaction between two hydrogen atoms. The temperature needed to ionize hydrogen atoms is of the order of $10^5$ K. At this temperature, the ratio (\ref{34}) can be estimated as $\frac{V(r)}{V_{N}(r)}\sim 10^{-36} \left(\frac{R}{r}\right)^{9}$, where we have taken $R$ as the Bohr radius ($\sim 10^{-11}$ m), and the transition wavelength $c/\omega_0\sim 10^{-7}$ m. Therefore, although it seems that when the temperature grows high enough, the interaction resulting from thermal fluctuations of gravitons may become appreciable to the classical Newtonian interaction.  Realistic physical objects, however, break up well before this temperature is  reached.

 \section{conclusion} 
 In this paper, we first introduce a relatively simple method to obtain the quadrupole-quadrupole  correction caused by  the quantum gravitational vacuum fluctuations to the classical Newtonian potential  between a pair of gravitationally polarizable objects. Then, we use it to treat the case where the gravitationally polarizable objects are in  a thermal bath of gravitons rather than in a vacuum. Assuming an approximately static polarizability  we find that in the high temperature limit the thermal fluctuations produce a dominant contribution and  the  potential  behaves like $r^{-10} T^{1}$. While, in the low temperature limit, the contribution from thermal fluctuations is proportional to $r^{-1}T^{10}$, which is much less than that from zero-point fluctuations, and in this case the  potential is dominated by the $r^{-11}$ term.

\acknowledgments  This work was supported by the National Natural Science Foundation of China under Grants No. 11435006, No. 11375092, No. 11447022, No. 11690034 and No. 11690030;  and the Zhejiang Provincial Natural Science Foundation of China under Grant No. LQ15A050001.

%%%%%%%%%%%%%%%%%%%%%%%%%%%%%%%%%%


\begin{thebibliography}{99}
\bibitem{Casimir}H. B. G. Casimir and D. Polder, Phys. Rev. {\bf73}, 360 (1948).
\bibitem{Lamoreaux05} S. K. Lamoreaux, Rep. Prog. Phys. {\bf 68}, 201 (2005).
\bibitem{Milton09} K. A. Milton, J. Phys.: Conf. Ser. {\bf 161}, 012001 (2009).
\bibitem{Bordag09} M. Bordag \textit{et al.}, \textit{Advances in the Casimir Effect} (Oxford University Press, Oxford, 2009).
\bibitem{Lifshitz}E. M. Lifshitz, Sov. Phys. JETP {\bf 2}, 73 (1956); I. E. Dzyaloshinskii, E. M. Lifshitz, and L. P. Pitaevskii, Adv. Phys. {\bf10}, 165 (1961); P. W. Milonni and A. Smith, Phys. Rev. A {\bf 53}, 3484 (1996); R. Passante and S. Spagnolo, Phys. Rev. A {\bf 76}, 042112  (2007).  

\bibitem{Ford16}L. H. Ford, M. P. Hertzberg, and J. Karouby, Phys. Rev. Lett. {\bf 116}, 151301 (2016).
\bibitem{Sernelius} B. E. Sernelius, {\it Surface modes in physics}, (Wiley-VCH, Berlin, 1999).

\bibitem{Donoghue}J. F. Donoghue, Phys. Rev. Lett. {\bf 72}, 2996 (1994); J. F. Donoghue, Phys. Rev. D {\bf50}, 3874 (1994); H. W. Hamber and S. Liu, Phys. Lett. B {\bf 357}, 51 (1995); I. B. Khriplovich and G. G. Kirilin, Zh. Eksp. Teor. Fiz. {\bf 95}, 1139 (2002) [J. Exp. Theor. Phys. {\bf 95}, 981 (2002)]; N. E. J. Bjerrum-Bohr, J. F. Donoghue, and B. R. Holstein, Phys. Rev. D{\bf 67}, 084033 (2003); {\bf71}, 069903 (2005).
 \bibitem{Wu16} P. Wu, J. Hu and H. Yu, Phys. Lett. B {\bf 763}, 40 (2016).
 \bibitem{Holstein2016}B. R. Holstein,  J. Phys. G {\bf 44},   01LT01 (2017);  arXiv:1610.07957.

\bibitem{Yu99}H. Yu and L. H. Ford, Phys. Rev. D {\bf 60},  084023 (1999).
\bibitem{Yu00}H. Yu and L. H. Ford, Phys. Lett. B {\bf 496}, 107 (2000).
\bibitem{Yu03}H. Yu and P. Wu,  Phys. Rev. D {\bf 68}, 084019  (2003). 
\bibitem{Yu09}H. Yu, N. F. Svaiter, and L. H. Ford, Phys. Rev. D {\bf 80}, 124019 (2009).

\bibitem{Oniga}T. Oniga and C. H.-T. Wang, Phys. Rev. D {\bf 94} , 061501(R) (2016).
\bibitem{Campbell}W. B. Campbell and T. A. Morgan, Am. J. Phys. {\bf 44}, 356 (1976); 
 A. Matte, Can. J. Math. {\bf 5}, 1 (1953);
 W. Campbell and T. Morgan, Physica (Amsterdam) {\bf 53}, 264 (1971);
P. Szekeres, Ann. Phys. (N.Y.) {\bf 64}, 599 (1971);
R. Maartens and B. A. Bassett, Classical Quant. Grav. {\bf 15}, 705 (1998);
 M. L. Ruggiero and A. Tartaglia, Nuovo Cimento B {\bf 117}, 743 (2002);
J. Ramos, M. de Montigny, and F. Khanna, Gen. Relativ. Gravit. {\bf 42}, 2403 (2010).


 \end{thebibliography}
\end{document}